\documentclass[10pt]{iopart}
\usepackage{iopams}

\usepackage{latexsym}            
\usepackage{psfig}
\usepackage{epsfig}
\usepackage{citesort}

\begin{document}

\newcommand{\beqn}{\begin{eqnarray}}
\newcommand{\eeqn}{\end{eqnarray}}

\title[Dynamics of an exclusion process...]{Dynamics of an
exclusion process with creation and annihilation}

\def\cH{{\cal H}}

\author{R\'obert Juh\'asz  and Ludger Santen}


\vskip 0.1cm
\address{ Theoretische Physik\\
Universit\"at des  Saarlandes\\
 66041 Saarbr\"ucken, Germany\\
juhasz@lusi.uni-sb.de, santen@lusi.uni-sb.de}

\begin{abstract}
We examine the dynamical properties of an exclusion process
with creation and annihilation  of particles in the framework of a
phenomenological 
domain-wall theory, by scaling arguments and by numerical simulation. 
We find that the length- and the
time scale are finite in the maximum current phase
for finite creation- and annihilation rates
as opposed to the algebraically decaying correlations
of the totally asymmetric simple exclusion process (TASEP).
Critical exponents of the transition to the TASEP
are determined.
The case where bulk creation- and annihilation rates vanish faster
than the inverse of the system size $N$ is also analyzed. 
We point out that shock localization
is possible even for rates proportional to  $N^{-a}$, $1< a<2$.
\end{abstract}


\submitto{\JPA}

\maketitle

\section{Introduction}

Self-driven many-particle systems have been extensively 
investigated recent years. The ongoing research interest in this kind 
of systems is both conceptual \cite{schutzrev,martinrev,dl,arndt,hinrichsen} 
and motivated by many important applications 
in different fields 
\cite{chowd,helbing_rmp,juelrmp,howbook,howard,alberts_book,leshouches,privman}.  
Stochastic models of self-driven many particle systems have been used in order
 to describe vehicular transport \cite{chowd,helbing_rmp}, pedestrian dynamics 
\cite{helbing_rmp}, 
intracellular transport \cite{juelrmp,howbook,howard} and many other problems
\cite{alberts_book,leshouches,privman}. 
The common feature of these models is the steady input of energy, 
which leads to generic non-equilibrium behaviour \cite{privman}. 
This feature implies that the standard methods applied in equilibrium statistical 
mechanics have to be generalized in order to handle such systems. 
First steps in this direction have been
made: Exact solutions of the stationary state have been 
obtained \cite{DEHP,SD}, a quantum formalism was established, 
 which rewrites the master equation as a Schr\"odinger equation
 in imaginary time \cite{schutzrev}, 
Yang-Lee zeroes have been introduced in order to describe non-equilibrium 
phase transition \cite{arndt} and for a number of transport models 
a free-energy formalism has been developed \cite{dl}. 

A question, which naturally arises, is how the 
properties of such systems are influenced by the presence or 
absence of conserved quantities. 
This issue has been studied by the example of  
the totally asymmetric simple exclusion process (TASEP)
\cite{tasep,krug}, which 
is the most simple non-trivial driven many particle model.
TASEP with particle reservoir which allow for particle exchange 
in the bulk was essentially introduced by Willmann {\it et al.}
as a model of a limit order market \cite{challet}. 
In this paper we study this model in the form as  defined by 
Parmeggiani {\it et al.} \cite{parm} therefore we refer to it as
PFF model in the following.  
This model can 
be viewed in the above sense as the grand-canonical 
counterpart of the TASEP and was motivated by the motion of molecular
motors,  which 
move along one-dimensional filaments 
\cite{juelrmp,alberts_book,howbook,howard}. It 
describes correctly the stochastic and biased motion of  particles, the
discrete structure of the filaments as well as the finite length 
of the path between attachment and detachment of a motor, which can be
tuned by adapting the capacity of the bulk reservoir. If one considers 
the PFF model with open boundaries the particle exchange
in the bulk may lead to the localization of an interface separating 
a high- and a low-density domain in the bulk. In addition to this, the
structure of the phase diagram differs strongly from that of the 
TASEP with particle conservation \cite{parm,ejs,popkov,LKN1,LKN2}.
In the present work we investigate the dynamical properties of this
model by phenomenological and numerical methods. 

The article is organized as follows. In the next section 
the model is re-introduced and the most important features of
its stationary state are reviewed. In section 3 we examine the dynamical
properties of the model in the shock phase 
by means of studying the density-density autocorrelation function which we
relate to a phenomenological theory of domain-wall motion. 
The phase diagram of the model is given in the case of  
vanishing total capacity of the bulk reservoir.
Phenomenological results are then compared to
direct simulations of the model.
In section 4 we discuss the length and time scales in the  
maximum current phase by means of scaling arguments and numerical simulation.
A summary and discussion of the results follows in
the final section.

\section{The model}

The PFF model \cite{parm,challet} is defined on a one-dimensional lattice
of $N$ sites,  each of which can either be empty
($\tau_i=0$) or occupied by a single  particle ($\tau_i=1$).
In the bulk of the system particles interact via asymmetric exclusion
dynamics, i.e. particle on site $i$ jump to the
neighbouring site $i+1$ with rate 1 provided it is empty.
Boundary sites are coupled to particle reservoirs, which realize
in- and output rates, whereas bulk sites change particles with a bulk
reservoir.
To be concrete: Particles enter
the system randomly on site 1 with attempt rate $\alpha$, and they can
leave it on the $L$th site with rate $\beta$.
At bulk sites $1,2,\dots ,L$ particles can attach with attempt rate
$\omega_A$ and detach with rate $\omega_D$.

The stationary density
$\langle \tau_i\rangle$ and current profiles $\langle j_i\rangle =
\langle \tau_i
(1-\tau_{i})\rangle$   have been recently studied by
a continuous mean field approximation \cite{parm,popkov,ejs}.
The attachment and detachment rates were taken to be
proportional to $1/N$, i.e. $\omega_A=\Omega_A/N$,
$\omega_D=\Omega_D/N$, where $\Omega_A$ and $\Omega_D$ are constants,
and the thermodynamical limit $N\to \infty$ was considered.
The lattice constant was rescaled according to $b=1/N$
so the spatial coordinate $x=i/N$ becomes a continuous variable in this limit.
Neglecting the
density-density correlations and the spatial
derivatives higher than the first-order one, one obtains the following equation
for the stationary density profile $\rho(x)$ \cite{ejs}:
\begin{equation}
(1-2\rho) \frac{ \partial \rho}{\partial x}
-\Omega_D\left[K -(1+K)\rho\right]=0,
\label{stst}
\end{equation}
where $K=\omega_A/\omega_D$.
The stationary density profile $\rho(x)$ can be constructed from
the flow-field of (\ref{stst}). In order to adapt the solution
of (\ref{stst}) to the boundary conditions one has to integrate
(\ref{stst}) from the left ($\rho(0)=\alpha$)  and right boundary
($\rho(1)=1-\beta$), respectively. This leads to the following
implicit expressions for the density profile \cite{ejs}
\beqn
x &=& \frac{1}{\Omega_D(1+K)}
 \frac{K-1}{(1+K)}
\ln \left| \frac{K-(1+K)\rho_-}{K-(1+K)\alpha}\right|
\nonumber \\
&+&\frac{2(\rho_--\alpha)}{\Omega_D(1+K)}
\nonumber \\
1-x &=& \frac{1}{\Omega_D(1+K)}  \frac{K-1}{(1+K)}
\ln \left| \frac{K-(1+K) \overline{\beta}}{K-(1+K)\rho_+}\right|
\\
&+& \frac{2(\overline{\beta} -\rho_+)}{\Omega_D(1+K)} \nonumber
\label{int}
\eeqn
where $\overline{\beta} = 1-\beta$ and $\rho_-$ ($\rho_+$) denotes the
 solution of (\ref{stst})
obtained from integration from the left (right) boundary.
The selection of the left or right solution is realized by means of
characteristics, which determine the velocity of discontinuity of
the density profile $\rho(x)$ \cite{ejs}. If the velocity of this
so-called shock is finite for any
position in the bulk $x$, it is driven out of the system and the stationary density profile is continuous.
Contrary to this,
a localized shock is observed if
\begin{equation}
\rho_-(x_s)+\rho_+(x_s)=1
\label{xs}
\end{equation}
holds for a particular position  $0<x_s<1$.
The profiles
constructed in the above way are believed to be exact in the $N\to \infty$ limit
\cite{popkov,ejs}.

In \cite{ejs} it has also been pointed out that the leading finite-size
corrections of the density profile  in the shock regime can be
obtained by applying the domain-wall theory for the dynamics of the
shock, which was originally developed for the TASEP \cite{ksks} and 
has been recently generalized to models without particle conservation
\cite{ejs,popkov}. The idea of this approach is to describe the stochastic
motion of the domain wall by a random walk with hopping rates determined
by the particle current in the low- and high-density domain. 
In the case of the TASEP the hopping rates are constant, as 
the current is constant due to particle conservation. 
Contrary to this, for
the PFF model one observes nontrivial current profiles $j(x)$, 
which lead to position-dependent hopping rates:
\begin{equation}
w_l(x) = \frac{j_-(x)}{\rho_+(x)-\rho_-(x)} \qquad
w_r(x) = \frac{j_+(x)}{\rho_+(x)-\rho_-(x)},
\label{hopping}
\end{equation}
where $\rho_{\pm}(x)$ and
$j_{\pm}(x)=\rho_{\pm}(x)(1-\rho_{\pm}(x))$ are the
density and the current in the high(+) and low(-) density domain,
respectively. 
The potential landscape governing the motion of the walker has a minimum at $x_s$, which we will refer to as equilibrium shock position.   
Previous analysis of the TASEP 
has shown that the random walk picture for the domain wall motion 
gives a correct description of time dependent
phenomena as well \cite{takesue,Dud,sanapp,nagy}. 
So we believe that this phenomenological description is appropriate also
for the dynamical properties of the PFF model.

\section{The shock phase}
\subsection{The case $\omega_{A,D}=\Omega_{A,D}/N$}

In this section we discuss the 
dynamical properties of the model for parameter combinations  where
the density profile has a discontinuity separating a  
high- and a low-density domain. 

First we consider the case where the total ``capacity'' of the bulk
reservoir is comparable with that of the 
boundary reservoirs,
i.e. $\omega_A=\Omega_A/N$, 
$\omega_D=\Omega_D/N$. The density profile of this model was
thoroughly studied and the parameter regime where the system exhibits
a shock is known (see the phase diagram in \cite{ejs}). 

In order to establish the relevant time scale, we  consider the
stationary (density-density) autocorrelation function
$C(i,t)\equiv \langle \tau_i(0)\tau_i(t)\rangle$,
where $\langle \dots \rangle$ denotes the average over the stationary
ensemble.
One expects that the dominant dynamical mode in a finite system,
which determines the long time behaviour of temporal correlations in the
vicinity of the equilibrium shock position $x_s$, is the stochastic
motion of the domain wall as opposed to ``microscopic'' processes.  
Thus, for large $N$ the local density at a given time $t$ is appropriately described by the function
\beqn
\tau(x,t) = \rho_+(x) + \left(\rho_-(x)-\rho_+(x) \right) \theta (\xi(t)-x)
\ ,
\label{rwdom}
\eeqn
where $\theta(x)$ is the Heaviside function and $\xi(t)$ is a random
walk with 
steps of length $1/N$  and with hopping rates given in
(\ref{hopping}).   
The potential well which the walker is trapped in is well approximated by
a harmonic potential in the vicinity of its minimum $x_s$. 
Considering the continuous description of the random walk the
Fokker-Planck equation reads \cite{kampen} 
\begin{equation}
{\partial P\over \partial t} = 
{V\over N}{\partial\over \partial y}yP(y,t)+
{D\over N^2}{\partial^2 P\over \partial y^2},
\label{fokker}
\end{equation}
where $y=x-x_s$ is the deviation from the equilibrium shock position. 
The constants $V$ and $D$, which characterize the shape of the 
potential, are given by 
\beqn
V & \equiv &  {d w_l(x_s)\over dx}-{d w_r(x_s)\over dx}, \nonumber \\  
D & \equiv & w_l(x_s)=w_r(x_s).
\label{const}
\eeqn
For the PFF model we have $V=\Omega_A+\Omega_D$. 
Equation (\ref{fokker}) is the
Fokker-Planck equation of the Ornstein-Uhlenbeck
process \cite{kampen}, and has the time-dependent solution:
\beqn 
P(y,t) &\equiv& P(y,t|y_0,0) \nonumber \\
&=& \left[{2\pi D\over N^2(\omega_A+\omega_D)}(1-e^{-2(\omega_A+\omega_D )t}) 
\right]^{-1/2} \nonumber \\
& \times & \exp \left[ -{N^2(\omega_A+\omega_D)\over
2D}{(y-y_0e^{-(\omega_A+\omega_D)t})^2\over
    1-e^{-2(\omega_A+\omega_D)t}}\right]  \ .
\label{sol}
\eeqn
For large system sizes the harmonic approximation is expected to give 
an appropriate description of the shock dynamics, since the localization
length of the walker increases only sub-extensively ($\sim N^{1/2}$). 

In the framework of the above phenomenological picture the autocorrelation function 
is given by
\beqn 
C(y,t)\equiv \langle \tau(y,0)\tau(y,t)\rangle \nonumber \\
 = \rho_-(y)[2\rho(y)-\rho_-(y)] + [\rho_+(y)-\rho_-(y)]^2I(y,t),
\label{corr1}
\eeqn
where
\beqn
I(y,t) \equiv \langle \theta(\xi(t)-y)\theta(\xi(0)-y)\rangle  \nonumber \\
= \int_{-x_s}^ydx_0\int_{-x_s}^ydx P(x+x_s,t|x_0+x_s,0)P_{st}(x_0+x_s). \qquad 
\label{I}
\eeqn
Inserting now (\ref{sol}) into (\ref{I}) we obtain 
\beqn 
I(y,t)=
\sqrt{{N^2(\omega_A+\omega_D) \over 8\pi D}}
\int_{-\infty}^0 e^{-{N^2(\omega_A+\omega_D)\over 2D}(x+y)^2} \nonumber \\
\times {\rm erfc}\left( \sqrt{ {N^2(\omega_A+\omega_D) \over 2D(1-T^2)}}
(T(x+y)-y)\right)dx,  
\label{I2}
\eeqn
where $T=e^{-(\omega_A+\omega_D)t}$ and erfc($x$) is the complementary
error function. 

The integral  $I(y,t)$ can be evaluated only at the equilibrium
position of the shock, 
i.e. for $y=0$, where we obtain the well-known result
\begin{equation} 
I(0,t)={1\over 4}+{1\over 2\pi}\arcsin [e^{-(\omega_A+\omega_D)t}].
\label{arcsin}
\end{equation}
For $y\neq 0$ $I(y,t)$ 
cannot be calculated analytically, however, it can be expanded for  short ($(\omega_A+\omega_D)t\ll 1$) and long times $(\omega_A+\omega_D)t\gg 1$). In the latter case the asymptotic form of $I(y,t)$ is given by: 

\beqn 
I(y,t)={1\over 4} \left(1+{\rm
erf}\left[y\sqrt{{N^2(\omega_A+\omega_D)\over 2D}}\right] \right)^2
\nonumber  \\
+{1\over 2\pi} e^{-{N^2(\omega_A+\omega_D)\over
D}y^2}e^{-(\omega_A+\omega_D)t} + {\mathcal{O}}(e^{-2(\omega_A+\omega_D)t}) \ .
\label{larget}
\eeqn
This expression shows that the motion 
of the domain wall introduces a time scale  $\tau={1\over
\omega_A+\omega_D}={N\over \Omega_A+\Omega_D}$, which is proportional to the system 
size. Note that this time scale $\tau$ is independent of the position
$y$. However, the domain wall contribution to
the true correlation function is relevant only in the region $|y|\ll N^{-1/2}$ and its amplitude is exponentially suppressed when leaving the equilibrium
shock position.

It is also interesting to discuss the short time behaviour of $I(y,t)$, i.e 
the case $(\omega_A+\omega_D)t\ll 1$. Here, the expansion of (\ref{I2}) yields
\beqn 
I(y,t)&=& {1\over 2} \left(1+{\rm
erf}\left[y\sqrt{{N^2(\omega_A+\omega_D)\over 2D}}\right] \right) 
\nonumber \\
&-& e^{-{N^2(\omega_A+\omega_D)\over 2D}y^2}{1\over
\pi}\sqrt{{\omega_A+\omega_D \over 2}}t^{1/2}+\dots 
\label{smallt}
\eeqn
The leading order correction is thus proportional to $t^{1/2}$, similarly to
the TASEP with parallel dynamics, where the space- and time-dependent
correlation function is exactly known \cite{schutz}.
This asymptotic form
corresponds
to a $f^{-3/2}$ power spectrum of the local density fluctuations as
it was found in the case of  the TASEP \cite{takesue}.

\subsection{The case of vanishing total capacity of the bulk reservoir}

We now turn to discuss the 
case where $\omega_A$ and $\omega_D$ vanish
faster than $1/N$.
In \cite{parm,popkov} it was then argued that the effect of bulk
reservoir is negligible in the thermodynamic limit, 
and the system will behave as the TASEP. 
We have found that this scenario is correct for any
parameter combination {\it except}  the line $\alpha=\beta< 1/2$.
Consider at this particular line the general case where attach and
detach rates scale as  $\omega_{A,D}=\Omega_{A,D}/N^{a}$.
We claim that a shock in
the density profile is still observed in the thermodynamic limit
whenever $1\le a<2$. 
Substituting $\omega_{A,D}=\Omega_{A,D}/N^{a}$  into the
calculations 
of the previous subsection one
obtains that the width of the shock grows with the system size as
$\xi=N\Delta x \sim N^{a\over 2}$,
whereas the time scale as
$\tau \sim N^{a}$.
Therefore, as long as  $1\le a<2$, the localization length
of the domain wall
increases only sub-extensively, and an unbounded motion
(as for the TASEP with $\alpha = \beta < 1/2$), is only
observed if $a>2$.
This is also true for the related time scale $\tau$. The time scale in
the TASEP is known to diverge proportionally to $N^2$
\cite{Dud,nagy,takesue}.
Therefore if $1\le a<2$ the time scale is determined by the attach and
detach processes ($\tau \sim N^{a}$), while TASEP-like behaviour ($\tau\sim N^2$) is observed
only if $a\ge 2$.

In the case $1<a<2$ the average density takes the values $\alpha$ in
the low-density domain and $1-\alpha$ in the high-density domain if
$N\to \infty$.
Thus, the location of the discontinuity in the $N\to\infty$ limit can be
obtained by substituting $\rho_-(x)={\rm const}=\alpha$ and
$\rho_+(x)={\rm const}=1-\alpha$ into (\ref{stst}). 
Equation (\ref{xs}) then yields
\begin{equation}
x_s={1-\alpha (1+K)\over (1+K)(1-2\alpha)}.
\end{equation}
Note that $x_s$ is independent of $a$.
Thus, a shock can be found in the system, whenever
$${\alpha\over 1-\alpha}<K<{1-\alpha\over \alpha}.$$
If ${\alpha\over 1-\alpha}>K$ ($K>{1-\alpha\over \alpha}$) the system
is in the (low-) high-density phase. 

For $\alpha\neq\beta$ and $a>1$ it is not possible to balance the current 
in the low- and high-density domain, i.e. to fulfill relation (\ref{xs}) 
for any $0<x<1$ in the $N\to\infty$ limit. Hence the shock
is driven out of the system and the density profile of the TASEP is 
recovered.

The description of the domain-wall dynamics by a random walk implies
that  
the dynamical exponent (defined as $\tau\sim\xi^z$) 
is $z=2$.
The same exponent can be observed for the TASEP at the
transition line
$\alpha = \beta<1/2$, where the 
length- and time scale are infinite 
in the limit $N\to\infty$, whereas for finite systems  
they scale as $\xi\sim N$ and
$\tau\sim N^2$, i.e. $z=2$. 
In the case of the TASEP this divergence is
restricted to the transition line $\alpha = \beta<1/2$, because the domain
wall motion is biased if $\alpha \neq \beta$.
In the case of the PFF model the position dependent
hopping rates introduce a localisation length which grows

sub-extensively in the whole shock phase.
Here, the
transition between shock- and low-density phase or
shock- and high-density phase
is not a localization-delocalization transition,
but it simply means that the equilibrium shock position moves to the
system boundaries. At the transition line the random
walker is trapped in a potential with hard-core repulsion for $y<0$
(HD-S transition) or $y>0$ (LD-S transition) respectively. The
complementary part of the potential landscape can still be described 
by the harmonic approximation.

Finally we mention that the domain wall contribution of the 
connected autocorrelation function 
$C_{conn}(y,t,N)\equiv\langle \tau(y,0)\tau(y,t)\rangle-
\langle \tau(y,0)\rangle^2$
has the following scaling form:
$$C_{conn}(y,t,N)=\tilde C(y^2N^{2-a},tN^{-a}).$$
We stress that 
the contribution to large time behaviour of the autocorrelation function, induced by the domain-wall
movement is relevant in a finite system only at sites $|y|\ll N^{{a\over 2}-1}$.

\subsection{Numerical results}

In this subsection we compare the phenomenological results
obtained in the previous parts of this section with
results of direct numerical simulation of the model.
In the numerical investigations we have mainly considered the case $a=1$,
 $\Omega_A=\Omega_D\equiv \Omega$,
where the equilibrium position of the shock $x_s$ and the diffusion constant
$D$ can be explicitly given. Equations (\ref{int}) can be solved for
$\rho_+(x)$ and $\rho_-(x)$ 
and condition (\ref{xs}) yields $x_s={\beta-\alpha \over 2 \Omega
}+{1\over 2}$.
The diffusion constant is given by 
$D={1\over 4}(1/\Delta-\Delta))$ where
$\Delta=1-\alpha-\beta-\Omega$ is the height of the shock \cite{ejs}.  
 
\begin{figure}
\epsfxsize=8truecm
\begin{center}
\mbox{\epsfbox{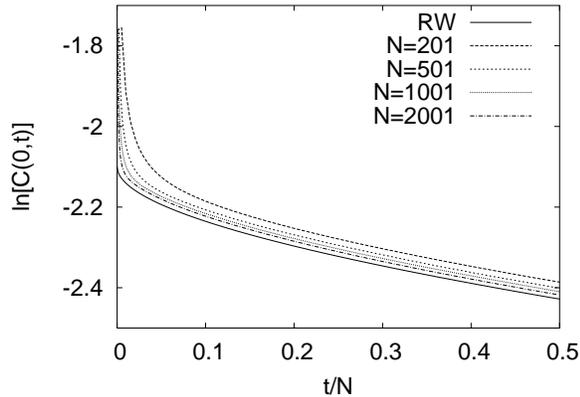}}
\end{center}
\caption{\label{fig1} The connected autocorrelation function
 measured at $y=0$ for different system sizes and with parameters
 $\alpha=0.1$,  $\beta=0.1$ and  $\Omega=0.1$. The solid line is
 the phenomenological prediction given by (\ref{corr1}) and
 (\ref{arcsin}).
}
\end{figure}

\begin{figure}
\epsfxsize=8truecm
\begin{center}
\mbox{\epsfbox{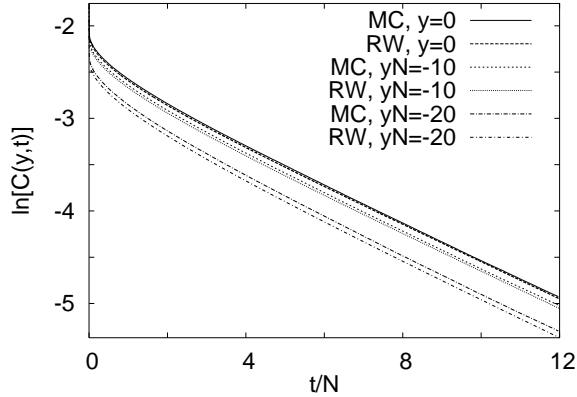}}
\end{center}
\caption{\label{fig2} The connected autocorrelation function
 measured at different positions in the chain for system size $N=1001$ and with parameters as in figure~\ref{fig1}. The
 phenomenological curve for $y\neq 0$  was obtained by integrating (\ref{I2}) numerically.
}
\end{figure}

The connected autocorrelation function was computed for system sizes $N=201-2001$. 
Results at the equilibrium position of the shock ($y=0$) are shown in
figure~\ref{fig1} and figure~\ref{fig2}. As can be seen in the figures
the phenomenological
predictions are in a good agreement with the numerical results except
for short times, and the accuracy of the phenomenological 
description improves for larger system sizes, as it was found in
the case of the stationary density profile \cite{ejs}.
A rapid decay of the autocorrelation function can be observed
on a time scale $t\sim {\mathcal{O}}(1)$,
which is connected to the ``self-correlation''  of particles,
i.e. the contribution of particles which are not updated during time $t$.
This ``microscopic'' time scale is related to the particle current and can be
identified as the finite time scale
in the high- and low density domains of the TASEP \cite{Dud,nagy}.
The above phenomenological picture apparently does not account for this
contribution, nevertheless the crossover to the regime,
which is dominated by the motion of the domain wall, takes place at rather short times
of  ${\mathcal{O}}(1/N)$.

In figure~\ref{fig2} results for sites away from the equilibrium shock
position ($y\neq 0$) are presented.
 The accuracy of the phenomenological theory is less satisfying
as leaving the equilibrium shock position.
This discrepancy, which increases with larger distances from the equilibrium shock position, may be related to the anharmonicity
of the potential-well.

\section{Maximum current phase}

The existence of a maximum current phase in the presence of bulk particle
exchange is restricted to  the case $\omega_A=\omega_D\equiv\omega$ (see
\cite{ejs}). 
Furthermore, we assume that the rate $\omega$ does not scale with the system size, i.e. $a=0$. 
Compared to the maximal current phase of the TASEP we
expect  that temporal correlations are reduced, as
the typical lifetime of particles is finite, and this
may introduce a finite time scale.
This was also observed in \cite{challet} where the correlation
function and fluctuations of price increments were investigated and the
corresponding time
scale was found to be $\omega^{-1}$.  
Here, we study the connected density-density autocorrelation 
function at the site in
 the middle of the chain.
Since the validity of the phenomenological treatment presented above
is restricted to the shock phase,
we resort to scaling arguments and numerical simulations.
To our knowledge this correlation function has not yet been
investigated in the maximum current phase even
in the case of particle conservation in the bulk. Therefore
we examine the dynamical correlations in the TASEP first.

\begin{figure}
\epsfxsize=8truecm
\begin{center}
\mbox{\epsfbox{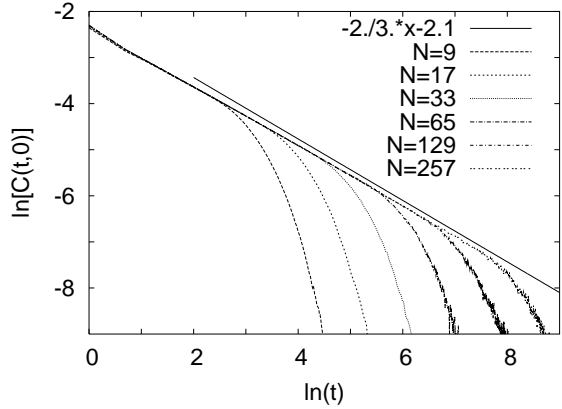}}
\end{center}
\caption{\label{figaut0} The connected autocorrelation function
 in the maximum current phase ($\alpha=\beta=0.5$) of the TASEP
 ($\omega_A=\omega_D=0$) for different system sizes.
}
\end{figure}

\begin{figure}
\epsfxsize=8truecm
\begin{center}
\mbox{\epsfbox{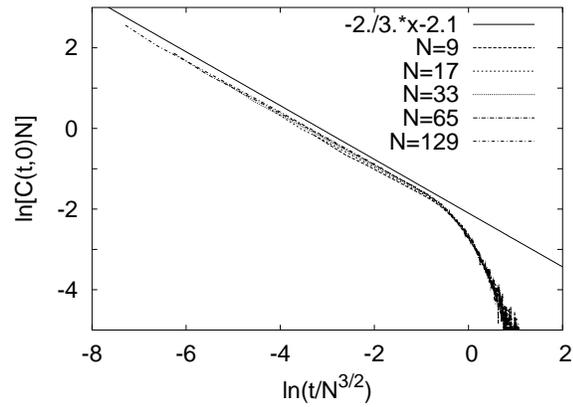}}
\end{center}
\caption{\label{figaut0+} The scaled connected autocorrelation function
 in the maximum current phase ($\alpha=\beta=0.5$) of the TASEP
 ($\omega_A=\omega_D=0$).
}
\end{figure}

Let us assume that the connected autocorrelation
function is a homogeneous function of its variables $N$ and $t$.
Rescaling then the lengths by a factor $b$ it transforms as
\begin{equation}
C(N,t)=b^{-x}\tilde C(N/b,t/b^{3/2}),  \qquad \omega =0
\label{homogen}
\end{equation}
where we have used that the dynamical exponent is $z=3/2$ in
the maximum current phase \cite{nagy,gwa}. The scaling dimension $x$
can be guessed as follows.  
The fluctuations of the total particle number scale with the
system size as $\sim N^{1/2}$ \cite{de,meakin}.
As a consequence the fluctuations of the local density
scale as  $\sim N^{-1/2}$. Since the autocorrelation function
contains a product of two local density operators, we have $x=1$.
Setting now $t=b^{3/2}$ in (\ref{homogen}) and taking  
the limit $N\to\infty$
we obtain $C(t)\sim t^{-2/3}$, whereas 
choosing $b=N$ we get 
\begin{equation}
C(N,t)=N^{-1}\Phi(t/N^{3/2}),
\label{phi}
\end{equation}
where the scaling function $\Phi(x)$ behaves as
\beqn
\Phi(x)\sim x^{-2/3} \qquad  x\ll 1 \nonumber \\
\Phi(x)\sim 0 \qquad         x\gg 1.
\label{scfunc}
\eeqn
Numerical results for the autocorrelation function are shown 
in figure~\ref{figaut0}, 
indicating an algebraic decay with an exponent
compatible with $2/3$, and the scaling plot in figure~\ref{figaut0+}
is in accordance with (\ref{phi}).

\begin{figure}
\epsfxsize=8truecm
\begin{center}
\mbox{\epsfbox{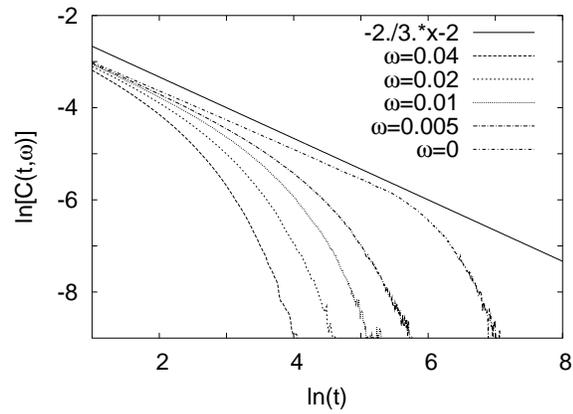}}
\end{center}
\caption{\label{figaut1} The connected autocorrelation function
 in the maximum current phase ($\alpha=\beta=0.5$) for different rates
 $\omega$. The size of the system is $N=257$.
}
\end{figure}

\begin{figure}
\epsfxsize=8truecm
\begin{center}
\mbox{\epsfbox{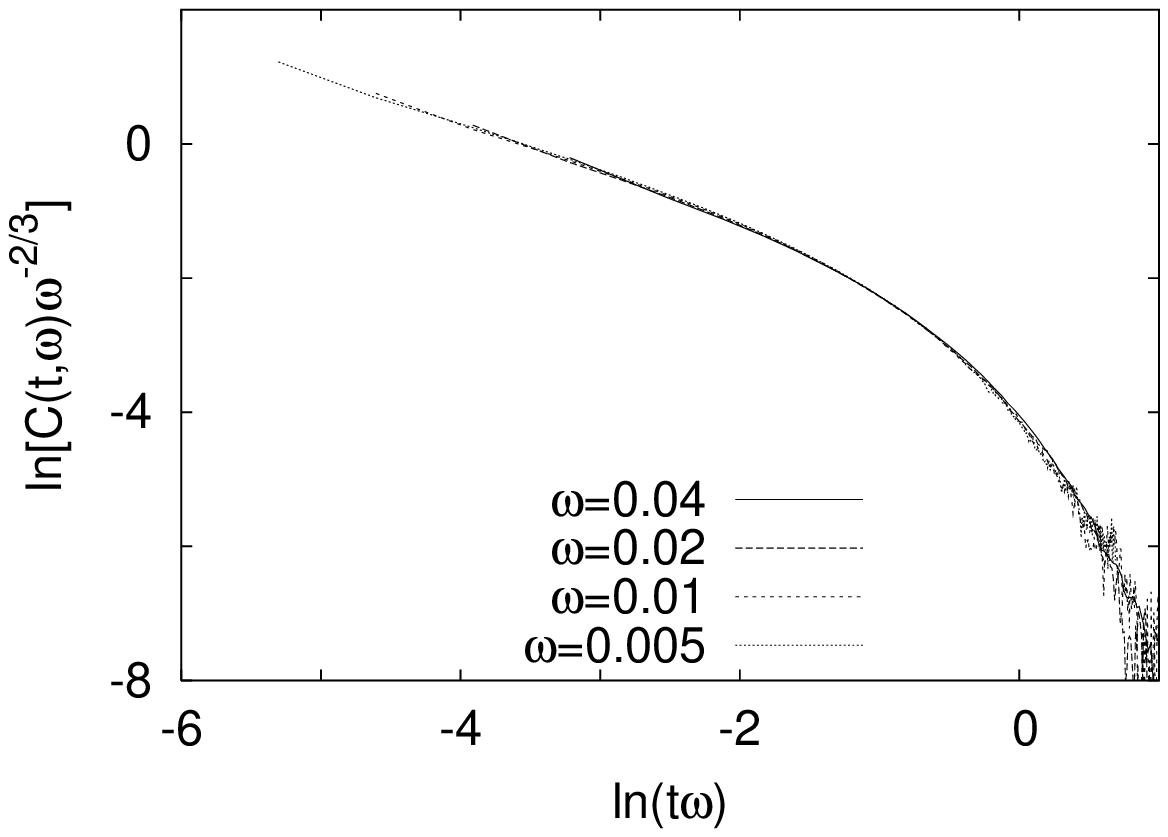}}
\end{center}
\caption{\label{figaut2} Scaling plot of the connected autocorrelation function
 in the maximum current phase ($\alpha=\beta=0.5$)
 according to (\ref{scaling_t}).}
\end{figure}

For finite $\omega$ and infinite system size one expects the following 
behaviour of the autocorrelation function when  time is rescaled by a
factor $b$: 
\begin{equation}
C(t,\omega)=b^{-2/3}\overline{C}(t/b,\omega b), \qquad 1/N=0.
\label{sc2}
\end{equation}
Setting $b=1/\omega$ we have 
\begin{equation}
C(t,\omega)=\omega^{2/3}\Psi(t\omega),  \qquad 1/N=0,
\label{scaling_t}
\end{equation}
where the scaling function $\Psi(x)$ behaves as $\Phi(x)$.
Numerical results
for finite $\omega$ are shown in figure~\ref{figaut1} and figure~\ref{figaut2}. 
These are in
agreement with (\ref{scaling_t}), showing a cut-off  at a
time scale $\tau \sim \omega^{-1}$
which is the typical lifetime of particles \cite{challet}.

Next, we determine the length scale $\xi$ in the infinite system.
In order to establish  $\xi$ we have studied the asymptotical
decay of the density profile.
For the TASEP the density profile is known to approach its bulk value $1/2$
algebraically \cite{krug,SD}.
Contrary to this, the particle creation and annihilation processes lead
to an exponentially fast asymptotical approach to the bulk density.
To see this,
consider the second-order mean field equation for the stationary
density profile \cite{parm} 
\begin{equation}
{1\over 2N}\frac{ \partial^2 \rho}{\partial x^2}+
(2\rho-1)({\partial \rho\over \partial x}-N\omega)=0.
\label{second}
\end{equation}
The  solution of (\ref{second}) in the asymptotic region where
$\rho(x)-{1\over 2}\ll 1$, has
the form $\rho(x)-{1\over 2}\sim e^{-Nx/\xi_{mf}}$. Putting this into
(\ref{second}) yields $\xi_{mf}={1\over 2\sqrt{\omega}}$. The scale of
the decay $ 
\xi_{mf}$ can then
be identified as a finite length scale of the system.
However, the true behaviour of $\xi$ is not recovered by the mean field
approximation as we show in the following.

\begin{figure}
\epsfxsize=8truecm
\begin{center}
\mbox{\epsfbox{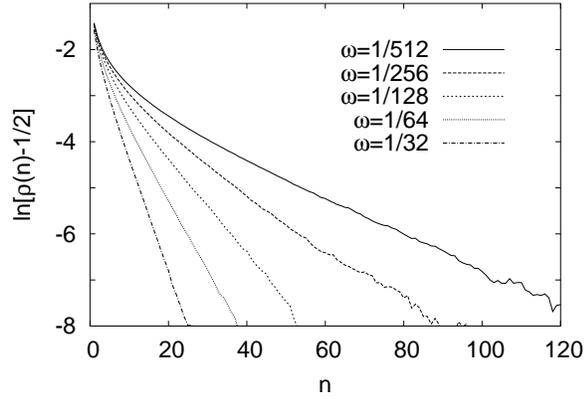}}
\end{center}
\caption{\label{figprof1} Density profile
 in the maximum current phase ($\alpha=\beta=1$) for different rates
 $\omega$. The size of the system 
 is N=512.}
\end{figure}

\begin{figure}
\epsfxsize=8truecm
\begin{center}
\mbox{\epsfbox{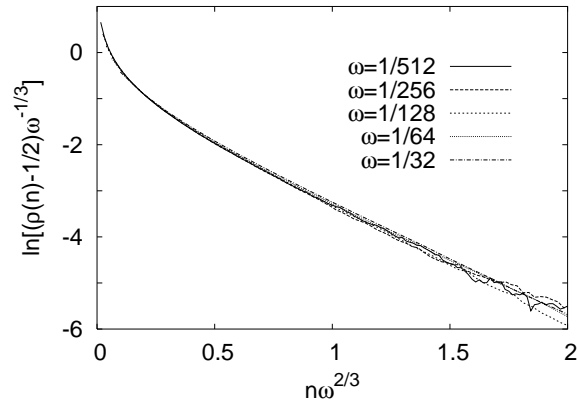}}
\end{center}
\caption{\label{figprof2} Scaling plot of the density profile 
 in the maximum current phase ($\alpha=\beta=1$). The size of the system
 is N=512.}
\end{figure}

Considering $\omega$ as a control parameter,
the point $\omega =0$ (TASEP) can be regarded as a critical point
of the PFF model, where length- and time scale diverge.
We expect
 that in the vicinity of the critical point, i.e. for the PFF model
 with $\omega \ll 1$ the dynamical exponent is the same as strictly at
 criticality, i.e. that $\tau\sim \xi^{3/2}$ holds. Comparing this with
 $\tau\sim \omega^{-1}$ we obtain 
\begin{equation}
\xi \sim \omega^{-\nu}, \qquad \nu=2/3.
\label{nu}
\end{equation}
 In this context the exponent $\nu$ plays the role of the correlation
length exponent. 
The correctness of (\ref{nu}) is
supported 
by numerical results for the density profile shown
in figure~\ref{figprof1}
and figure~\ref{figprof2}. 

Thus, we have seen that the length- and the time scales remain finite in the
thermodynamic limit as long as $\omega$ does not vanish in that limit,
i.e. $a=0$. If $a>0$, apparently both the length- and the time scale diverge in the
$N\to \infty$ limit, as in the case of the TASEP.
However, for $a<{3\over 2}$  they scale  with
exponents that are different from those of the TASEP, viz. $\xi\sim
N^{{2\over 3}c}$,  $\tau\sim N^c$, 
where $c=\min\{a,{3\over 2}\}$.


\section{Discussion}

The analysis of the dynamical properties completes the description
of the stationary state of the PFF model. In the framework of a
phenomenological domain-wall theory we could establish  a time scale
in the vicinity of 
the equilibrium domain wall position,  which 
grows with
the system size as $\sim N^{a}$. 
The localisation length of the domain-wall 
scales as
$\sim N^{a\over 2}$, leading to the dynamical exponent $z=2$,
which is identical to that of the TASEP at the coexistence line
$\alpha = \beta$. 
This is in both cases the consequence of the diffusive nature of the
dominant dynamical mode. However, in the 
PFF model it is relevant
only in a subextensively growing region whereas in the latter one in
the whole system. 
Aside from the shock region the time scale is finite,
and is related to the inverse of the particle current.
Contrary to the TASEP the
current is position dependent, a feature of the model which is reflected
by the position dependence of the time scale.
This position-dependent microscopic time scale is observed
in the high- and low-density phases, as well.
We note that the transition from the shock- to high-density
(or low-density) phase manifests itself simply in the equilibrium
domain-wall position leaving the system
and it is not a delocalisation transition 
as in the case
of the TASEP.
We have pointed out that for $\alpha =\beta<1/2$ even a 
vanishing total capacity of
the bulk reservoir ($1<a<2$) is 
able to localise the shock.
We mention that the phenomenological calculations presented in
this work can be carried out for arbitrary systems possessing a
localised fluctuating domain wall, such as the model recently
introduced by R\'akos 
{\it et al.} \cite{rakos}.

In the maximum current phase of the TASEP 
we have found the algebraic decay of the
autocorrelation function. 
By scaling arguments we could determine
the decay exponent which was found to be $2/3$.
The introduction of noise through bulk particle exchange 
with a finite rate
destroys
the power-law correlations and the resulting phase is
characterised by finite length- and time scales.
Thus, the TASEP can be regarded as a critical point of the PFF model.
By scaling arguments we have determined the critical exponents which are
in accordance with the results of numerical simulations.

\ack

We thank F. Igl\'oi and J. Krug for useful discussions. 
This work has been supported by the Deutsche Forschungsgemeinschaft
under Grant No. SA864/2-1. \\

\bibliographystyle{unsrt}

\end{document}